\documentclass[fleqn,usenatbib]{mnras}

\usepackage{newtxtext,newtxmath}

\usepackage[T1]{fontenc}

\DeclareRobustCommand{\VAN}[3]{#2}
\let\VANthebibliography\thebibliography
\def\thebibliography{\DeclareRobustCommand{\VAN}[3]{##3}\VANthebibliography}

\usepackage{graphicx}	
\usepackage{amsmath}	

\newcommand{\Msun}{\ensuremath{\mathrm{M}_\odot}}
\newcommand{\Msunpyr}{\ensuremath{\Msun~\mathrm{yr^{-1}}}}
\newcommand{\kmps}{\ensuremath{\mathrm{km~s^{-1}}}}

\graphicspath{{./}{figures/}}

\title[Slowly rising interacting SNe]{On the nature of slowly rising interaction-powered supernovae}

\author[T. J. Moriya]{
Takashi J. Moriya,$^{1,2}$\thanks{E-mail: takashi.moriya@nao.ac.jp (TJM)}
\\
$^{1}$National Astronomical Observatory of Japan, National Institutes of Natural Sciences, 2-21-1 Osawa, Mitaka, Tokyo 181-8588, Japan \\
$^{2}$School of Physics and Astronomy, Faculty of Science, Monash University, Clayton, Victoria 3800, Australia
}

\date{Accepted 2023 July 19. Received 2023 June 22; in original form 2023 May 22}

\pubyear{2023}

\begin{document}
\label{firstpage}
\pagerange{\pageref{firstpage}--\pageref{lastpage}}
\maketitle

\begin{abstract}
Some interaction-powered supernovae have long rise times of more than 100~days. We show that such long rise times are naturally expected if circumstellar matters (CSM) have a flat density structure ($s\lesssim 1.5$, where $\rho_\mathrm{CSM}\propto r^{-s}$). In such cases, bolometric luminosities from the CSM interaction keep increasing as long as the CSM interacts with the outer layers of the SN ejecta. Thus, the rise time is determined by the dynamical timescale in which the reverse shock propagates the outer layers of the SN ejecta, not by the timescales in which photons diffuse in the CSM as often considered. Interaction-powered supernovae with very long rise times can be an important probe of extensive non-steady mass loss in massive stars.
\end{abstract}




\section{Introduction}
Core-collapse supernovae (SNe) are explosions of massive stars. Massive stars lose their mass throughout their evolution \cite[e.g.,][]{vink2022}, and core-collapse SNe occur within a circumstellar matter (CSM) formed by their progenitors. Especially, Type~IIn, Ibn, and Icn SNe (SNe~IIn, Ibn, Icn) are known to show strong signatures of the interaction between SN ejecta and CSM, and their luminosities and spectra are dominated by the interaction signatures (see \citealt{fraser2020} for a recent review). The progenitors of such interaction-powered SNe are estimated to have mass-loss rates exceeding $10^{-4}~\Msunpyr$ which are much higher than those typically observed in massive stars \citep[e.g.,][]{fox2011,kiewe2012,taddia2013,ofek2014,maeda2022}. Light curves and spectra of interaction-powered SNe contain rich information on the mysterious extensive mass loss of massive stars \citep[e.g.,][]{moriya2013,moriya2014,dessart2016}. The CSM interaction may also play an important role in providing luminosities of some superluminous SNe, even if there are no apparent spectral signatures of the CSM interaction \citep[e.g.,][]{moriya2012,mazzali2016,chen2017,wheeler2017}.

Some interaction-powered SNe are known to have very long rise times. For example, HSC16aayt is a SN~IIn with a rise time of more than 100~days \citep{moriya2019,moriya2021}. SN~2008iy is a similar, slightly less luminous SN~IIn with a rise time of a few hundred days \citep{miller2010}. The origin of such a long rise time lasting for more than 100~days is not clear. The rise times of interaction-powered SNe are usually considered to correspond to the diffusion time of photons in the CSM \citep[e.g.,][]{smith2007,chevalier2011}. However, in order to realize the diffusion time of more than 100~days, it is usually found that a dense CSM with around 100~\Msun\ or more is required \citep[e.g.,][]{suzuki2020,khatami2023}. Such a massive CSM exceeding 100~\Msun\ is difficult to be formed.

In this paper, we present that interaction-powered SNe are naturally expected to have long rise times exceeding 100~days if the CSM density structure is much flatter than that expected from steady mass loss. In such a case, the rise time can be more than 100~days because of dynamics even if the diffusion time is much less than 100~days. First, we introduce our formulation of interaction-powered SN light curves in Section~\ref{sec:lightcurvemodel}. We compare our theoretical light curves with observed long-rising SNe~IIn in Section~\ref{sec:comparison}. We discuss and summarize our findings in this paper in Section~\ref{sec:summary}.

\section{Bolometric light curves of interaction-powered supernovae}\label{sec:lightcurvemodel}
We adopt the same formalism to estimate the luminosity evolution of interacting SNe as in previous studies \citep[e.g.,][]{chatzopoulos2012,moriya2013,hiramatsu2023}. We provide a brief summary of the bolometric light-curve model.

We consider SNe in which luminosity is dominated by the interaction between SN ejecta and dense CSM. In such a case, a strong radiative cooling in the shocked region is expected, and the shocked region becomes a cool, dense, thin shell \citep[e.g.,][]{chugai2004,vanmarle2010}. Thus, we assume that the thickness of the shocked region ($\Delta r_\mathrm{sh}$) is much smaller than the shock radius ($r_\mathrm{sh}$), i.e., $\Delta r_\mathrm{sh}\ll r_\mathrm{sh}$. Assuming a spherical symmetry, the evolution of $r_\mathrm{sh}$ in such a case can be evaluated by the following momentum equation,
\begin{equation}
    M_\mathrm{sh}\frac{dv_\mathrm{sh}}{dt} = 4\pi r_\mathrm{sh}^2\left[\rho_\mathrm{ej}(v_\mathrm{ej}-v_\mathrm{sh})^2 - \rho_\mathrm{CSM}(v_\mathrm{sh}-v_\mathrm{CSM})^2 \right], \label{eq:momentum}
\end{equation}
where $v_\mathrm{sh}=dr_\mathrm{sh}/dt$ is the shocked shell velocity, $\rho_\mathrm{ej}$ is the ejecta density, $v_\mathrm{ej}$ is the ejecta velocity, $\rho_\mathrm{CSM}$ is the CSM density, and $v_\mathrm{CSM}$ is the CSM velocity. We assume a single power-law density structure for the CSM ($\rho_\mathrm{CSM} = D r^{-s}$) and a constant CSM velocity.
Depending on the unknown mass-loss mechanisms forming the dense CSM surrounding interaction-powered SNe, the dense CSM may not have a smooth density structure as assumed here \citep[e.g.,][]{smith2014}. However, it is generally found that light curves of interaction-powered SNe roughly follow a power-law evolution expected from a smooth density structure \citep[e.g.,][]{fransson2014}. In addition, only 1-10\% of SNe~IIn are known to have conspicuous light-curve bumps that would appear if the dense CSM has a bumpy structure, for example \cite[e.g.,][]{nyholm2020}. Thus, our assumption of a smooth CSM is reasonable enough to obtain a rough estimate of the CSM and explosion properties in interaction-powered SNe.

SN ejecta are assumed to have a two-component power-law density structure, i.e., $\rho_\mathrm{ej}\propto r^{-n}$ outside and $\rho_\mathrm{ej}\propto r^{-\delta}$ inside \citep[e.g.,][]{chevailer1989}. We assume a homologous expansion of the SN ejecta ($v_\mathrm{ej} = r/t$), and thus we can express $\rho_\mathrm{ej}$ as
\begin{equation}
\rho_\mathrm{ej}(v_\mathrm{ej},t)=\left\{ \begin{array}{ll}
\frac{1}{4\pi(n-\delta)}
\frac{[2(5-\delta)(n-5)E_\mathrm{ej}]^{(n-3)/2}}{
[(3-\delta)(n-3)M_\mathrm{ej}]^{(n-5)/2}}
t^{-3}v_\mathrm{ej}^{-n} & (v_\mathrm{ej}>v_t), \\ 
\frac{1}{4\pi(n-\delta)}
\frac{[2(5-\delta)(n-5)E_\mathrm{ej}]^{(\delta-3)/2}}{
[(3-\delta)(n-3)M_\mathrm{ej}]^{(\delta-5)/2}}
t^{-3}v_\mathrm{ej}^{-\delta} &
 (v_\mathrm{ej}<v_t), \\ 
\end{array} \right.
\end{equation}
where $E_\mathrm{ej}$ and $M_\mathrm{ej}$ are kinetic energy and mass of the SN ejecta, respectively. $v_t$ is the velocity in the SN ejecta where the inner and outer density structure is connected, i.e.,
\begin{equation}
    v_t=\left[\frac{2(5-\delta)(n-5)E_\mathrm{ej}}{(3-\delta)(n-3)M_\mathrm{ej}}\right]^{\frac{1}{2}}.
\end{equation}
Numerical simulations of stellar explosions indicate that $n=7-12$ and $\delta=0-1$, depending on progenitors and explosion mechanisms \citep[e.g.,][]{matzner1999}. 

Shortly after a SN explosion, the outer SN ejecta ($\rho_\mathrm{ej}\propto r^{-n}$) interact with the surrounding CSM ($\rho_\mathrm{CSM}=Dr^{-s}$). In this phase, Eq.~(\ref{eq:momentum}) provides an analytic solution for $r_\mathrm{sh}$ which is expressed as
\begin{equation}
    r_\mathrm{sh} (t)= A t^{\frac{n-3}{n-s}}, \label{eq:rsh}
\end{equation}
where
\begin{equation}
A = \left[
\frac{(3-s)(4-s)[2(5-\delta)(n-5)E_\mathrm{ej}]^{\frac{n-3}{2}}}{4\pi D(n-4)(n-3)(n-\delta)[(3-\delta)(n-3)M_\mathrm{ej}]^{\frac{n-5}{2}}}
\right]^{\frac{1}{n-s}}.
\end{equation}
Assuming that a fraction $\varepsilon$ of kinetic energy coming into the shocked shell is converted to radiation, we can express the bolometric luminosity of an interaction-powered SN as
\begin{equation}
    L = 2\pi \varepsilon \rho_\mathrm{CSM} r_\mathrm{sh}^2v_\mathrm{sh}^3,
\end{equation}
where $\varepsilon$ is a conversion efficiency from kinetic energy to radiation. We assume $\varepsilon=0.3$ in this work \citep[e.g.,][]{fransson2014}. Given Eq.~(\ref{eq:rsh}), the luminosity evolution can be expressed as 
\begin{equation}
    L = L_1 t^{\alpha}, \label{eq:luminosity}
\end{equation}
where
\begin{eqnarray}
    L_1 &=& \frac{\varepsilon}{2}\left(4\pi D\right)^{\frac{n-5}{n-s}}
\left(\frac{n-3}{n-s}\right)^3 \nonumber \\
&& \times
\left[
\frac{(3-s)(4-s)[2(5-\delta)(n-5)E_\mathrm{ej}]^{\frac{n-3}{2}}}{(n-4)(n-3)(n-\delta)[(3-\delta)(n-3)M_\mathrm{ej}]^{\frac{n-5}{2}}}
\right]^{\frac{5-s}{n-s}},
\end{eqnarray}
and
\begin{equation}
    \alpha = \frac{6s-15+2n-ns}{n-s}. \label{eq:alpha}
\end{equation}
The bolometric light-curve evolution expressed by Eq.~(\ref{eq:luminosity}) is valid as long as the SN ejecta density structure entering the shocked region through the reverse shock follows $\rho_\mathrm{ej}\propto r^{-n}$. The reverse shock reaches the transition velocity $v_t$ at
\begin{equation}
t_t=\left[\frac{(3-s)(4-s)}{4\pi D(n-4)(n-3)(n-\delta)}
\frac{[(3-\delta)(n-3)M_\mathrm{ej}]^{\frac{5-s}{2}}}{[2(5-\delta)(n-5)E_\mathrm{ej}]^{\frac{3-s}{2}}}
\right]^{\frac{1}{3-s}}. \label{eq:tt}
\end{equation}
After $t_t$, the inner SN ejecta with the density structure $\rho_\mathrm{ej}\propto r^{-\delta}$ start to enter the shocked region. We assume $\delta=0$ in this work.

\begin{figure}
	\includegraphics[width=\columnwidth]{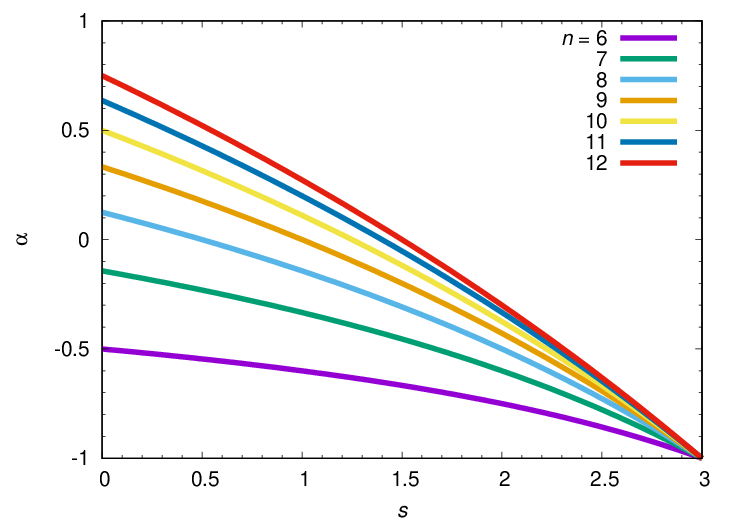}
	\includegraphics[width=\columnwidth]{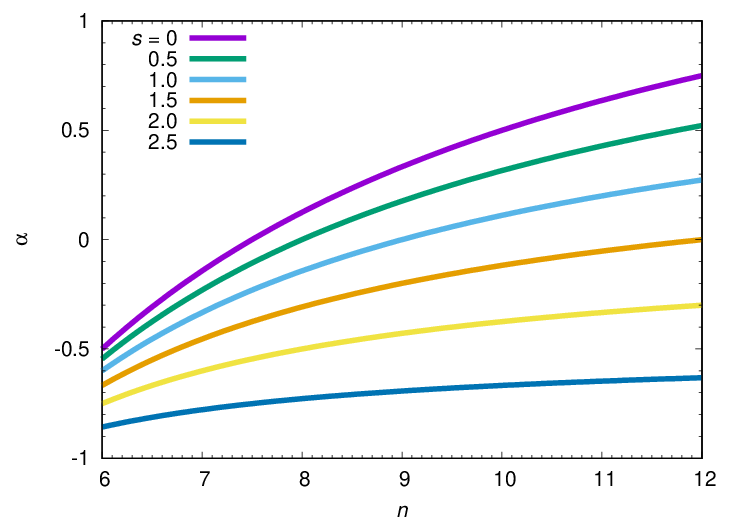}
    \caption{
    Power ($\alpha=[6s-15+2n-ns]/[n-s]$) of bolometric light curves ($L\propto t^{\alpha}$) of interaction-powered SNe in early phases before $t_t$ from the analytic model. The top panel shows $\alpha$ as a function of $s$ for different $n$, and the bottom panel shows $\alpha$ as a function of $n$ for different $s$.
    }
    \label{fig:alpha}
\end{figure}

\begin{figure}
	\includegraphics[width=\columnwidth]{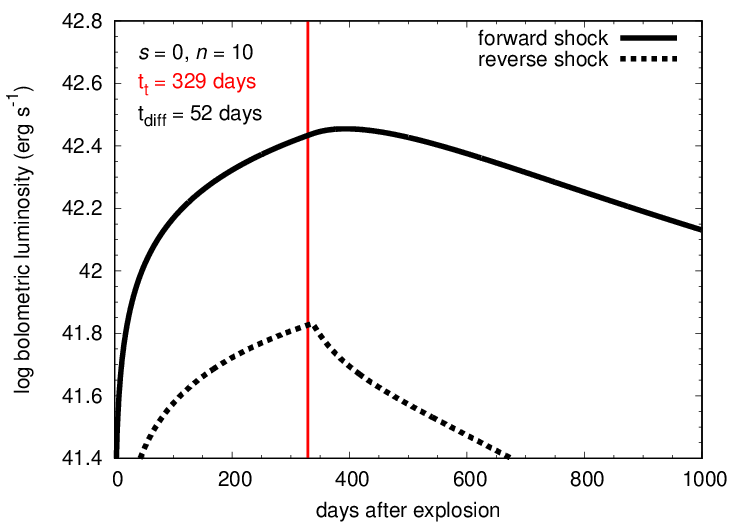}
	\includegraphics[width=\columnwidth]{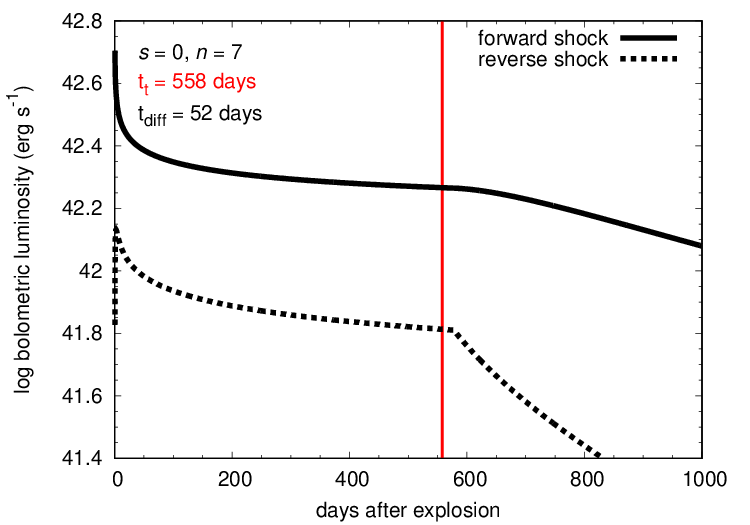}
    \caption{
    Representative interaction-powered SN light curve models for the flat CSM density ($s=0$). The CSM density is $\rho_\mathrm{CSM}=10^{-15}~\mathrm{g~cm^{-3}}$. The top panel shows the model with $n=10$ and the bottom panel shows the model with $n=7$. The SN ejecta mass of $M_\mathrm{ej}=10~\Msun$, the explosion energy of $E_\mathrm{ej}=10^{51}~\mathrm{erg}$, and $\delta=0$ are assumed. The solid lines show luminosities from the forward shock (Eq.~\ref{eq:luminosity}) and the dashed lines show luminosities from the reverse shock. The transitional time $t_t$ (Eq.~\ref{eq:tt}) and diffusion time $t_\mathrm{diff}$ (Eq.~\ref{eq:tdiff}) are shown.
    }
    \label{fig:lightcurve_s0}
\end{figure}

Figure~\ref{fig:alpha} shows the power-law index $\alpha$ for the bolometric luminosity evolution of interaction-powered SNe (Eqs.~\ref{eq:luminosity} and \ref{eq:alpha}). When mass loss from progenitors is close to steady $(s\simeq 2)$, we can find that $\alpha$ is always negative in the parameter range we are interested in ($7\lesssim n\lesssim 12$). In such a case, after the initial luminosity rise caused by a photon diffusion in a dense CSM, interaction-powered SN light curves are predicted to decline regardless of the SN ejecta density structure. However, if the CSM density structure is flatter, we can find that $\alpha$ can be positive, i.e., luminosity can increase with time. For example, when $s=0$, $\alpha$ is positive if $n>7.5$. In such a case, interaction-powered light curves keep rising as long as the SN ejecta structure entering the reverse shock follows $\rho_\mathrm{ej}\propto r^{-n}$ ($t<t_t$). After the inner SN ejecta with $\rho_\mathrm{ej}\propto r^{-\delta}$ start to interact ($t>t_t$), the interaction-powered light curves are expected to start declining gradually. $\alpha$ can be positive when $s\lesssim 1.5$ within the range of $n$ we are interested in. 

We have discussed an analytic solution to interaction-powered light curves so far. Equation~(\ref{eq:momentum}) can be solved numerically, and the whole light curve evolution can be evaluated through the numerical solution. Figure~\ref{fig:lightcurve_s0} shows examples of estimated light-curve evolution with $s=0$ for $n=10$ and $7$. In these examples, SN ejecta are assumed to have $M_\mathrm{ej}=10~\Msun$ and $E_\mathrm{ej}=10^{51}~\mathrm{erg}\equiv1~\mathrm{B}$. The CSM density is set as $\rho_\mathrm{CSM}=10^{-15}~\mathrm{g~cm^{-3}}$ and the CSM velocity is assumed to be $v_\mathrm{CSM}=100~\kmps$. When $n=10$, the light curve slowly rises at first as expected by the analytic solution. The light curve starts to decline shortly after $t_t=329~\mathrm{days}$. 

The diffusion time ($t_\mathrm{diff}$) in CSM is an important parameter that characterizes the light curves of interaction-powered SNe. The diffusion time can be expressed as
\begin{equation}
    t_\mathrm{diff} = \frac{\tau_\mathrm{CSM}R_\mathrm{CSM}}{c}, \label{eq:tdiff}
\end{equation}
where $\tau_\mathrm{CSM}$ is the optical depth in the CSM, $R_\mathrm{CSM}$ is the CSM radius, and $c$ is the speed of light. In an example discussed in Fig.~\ref{fig:lightcurve_s0}, $r_\mathrm{sh}$ reaches $\simeq 2\times 10^{16}~\mathrm{cm}$ at 1000~days, and we set $R_\mathrm{CSM}=2\times 10^{16}~\mathrm{cm}$. Assuming that the CSM opacity $\kappa_\mathrm{CSM}$ is dominated by the electron scattering opacity and the CSM has the solar composition, $\kappa_\mathrm{CSM}=0.34~\mathrm{cm^{2}~g^{-1}}$ when it is fully ionized. Thus, the diffusion time in the CSM in the light-curve model in Fig.~\ref{fig:lightcurve_s0} is $t_\mathrm{diff}=52~\mathrm{days}$. The diffusion time is shorter than the transitional time $t_t=329~\mathrm{days}$, and the light-curve rise time is determined by $t_t$, not $t_\mathrm{diff}$. The rise time of around 400~days is realized by the CSM mass of only 17~\Msun.

\begin{figure}
	\includegraphics[width=\columnwidth]{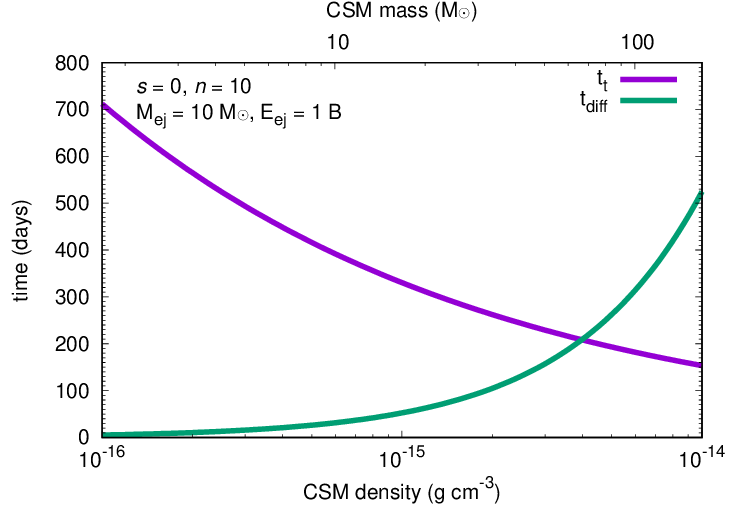}
    \caption{
    Comparison between the transitional time $t_t$ (Eq.~\ref{eq:tt}) and the diffusion time (Eq.~\ref{eq:tdiff}) for the case of the flat CSM density ($s=0$) with $n=10$. The CSM radius is assumed to be $R_\mathrm{CSM}=2\times 10^{16}~\mathrm{cm}$. $t_t$ and $t_\mathrm{diff}$ are presented as a function of the CSM density, and the corresponding CSM mass is also shown. The SN ejecta mass of $M_\mathrm{ej}=10~\Msun$, the explosion energy of $E_\mathrm{ej}=10^{51}~\mathrm{erg}$, and $\delta=0$ are assumed.
    }
    \label{fig:tt}
\end{figure}

An example of the comparison between the transitional time $t_t$ and the diffusion time $t_\mathrm{diff}$ for $s=0$, $n=10$, $M_\mathrm{ej}=10~\Msun$, $E_\mathrm{ej}=1~\mathrm{B}$, and $R_\mathrm{CSM}=2\times 10^{16}~\mathrm{cm}$ is presented in Fig.~\ref{fig:tt}. When the CSM density is low, $t_t$ is larger than $t_\mathrm{diff}$ and the light curve rise time is determined by $t_t$. The light curve rise time can be several hundred days even if the diffusion time is much smaller. The diffusion time becomes larger than the transitional time when $\rho_\mathrm{CSM}\gtrsim 4\times 10^{-15}~\mathrm{g~cm^{-3}}$ where the CSM mass becomes larger than $67~\Msun$. Our result shown that the rise time of several hundred days can be achieved even if the CSM mass is $\sim 1-10~\Msun$ if the CSM density is flat.

Equation~(\ref{eq:luminosity}) assumes that the interaction luminosity from the forward shock entering the CSM dominates. In Fig.~\ref{fig:lightcurve_s0}, we show the luminosities from both forward and reverse shocks. We can find that the luminosity from the forward shock dominates the interaction-powered SNe.

\begin{figure}
	\includegraphics[width=\columnwidth]{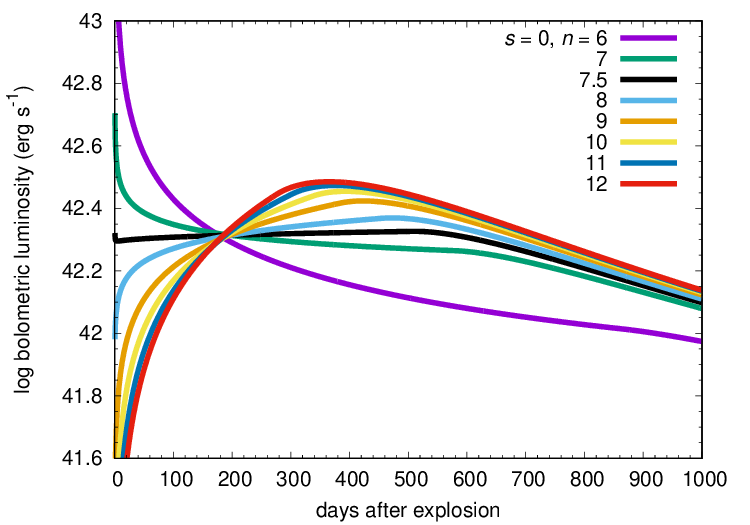}
	\includegraphics[width=\columnwidth]{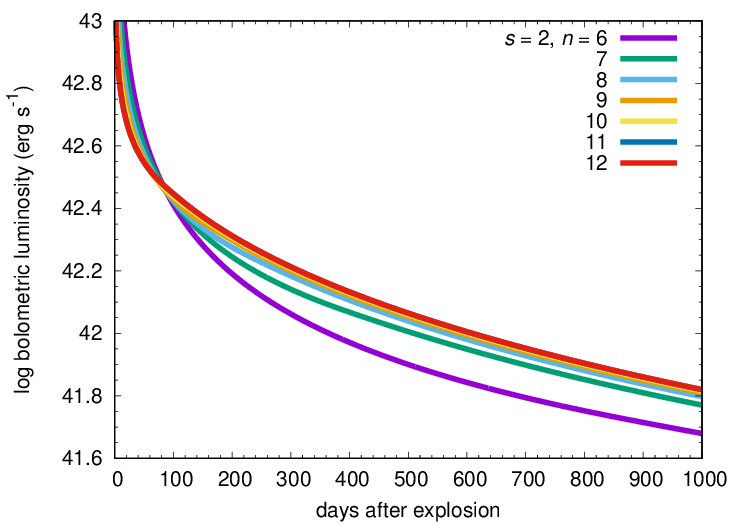}
    \caption{
    Numerical bolometric light curve models of interaction-powered SNe for $s=0$ (top) and $s=2$ (bottom). The assumed CSM density is $\rho_\mathrm{CSM} = 10^{-15}~\mathrm{g~cm^{-3}}$ (top) and $\rho_\mathrm{CSM}=5\times 10^{-16}(r/10^{16~\mathrm{cm}})^{-2}~\mathrm{g~cm^{-3}}$ (bottom). The effects of the ejecta density structure ($n$) on the light curves are presented. The SN ejecta mass of $M_\mathrm{ej}=10~\Msun$, the explosion energy of $E_\mathrm{ej}=10^{51}~\mathrm{erg}$, and $\delta=0$ are assumed. 
    }
    \label{fig:lightcurve_s}
\end{figure}

Figure~\ref{fig:lightcurve_s0} presents a light curve model for $n=7$ for which $\alpha$ becomes negative. The light curve keep declining from the beginning as expected by the analytic solution. After $t_t$, the light curve starts to decline faster. A series of numerical light curve models for $s=0$ and $s=2$ with different $n$ are presented in Fig.~\ref{fig:lightcurve_s}. As discussed so far, we can find slowly rising light curves can be found in some $s=0$ models, while all the light curve models with $s=2$ simply decline.


\begin{table*}
	\centering
	\caption{Estimated parameters for HSC16aayt and SN~2008iy}
	\label{tab:models}
	\begin{tabular}{lccccccccc} 
		\hline
		SN name & $n$ & $\delta$ & $E_\mathrm{ej}$ & $M_\mathrm{ej}$ & $\rho_\mathrm{CSM}$ & $M_\mathrm{CSM}$ & $t_t$ & $t_\mathrm{diff}$ \\
		     & &  & $(\mathrm{B})$ & $(\Msun)$ & $(\mathrm{g~cm^{-3}})$ & $(\Msun)$ & (days) & (days) \\  
		\hline
		HSC16aayt  & $10$ & $0$ & $5$ & $10$ & $10^{-15}$ & $29~\Msun^a$ & $147$ & $76^a$ \\
		SN~2008iy  & $10$ & $0$ & $2.5$ & $3$ & $10^{-16}$ & $8~\Msun^b$ & $164$ & $14^b$ \\
		\hline
    \multicolumn{9}{l}{$^a$ $R_\mathrm{CSM}=2.4\times 10^{16}~\mathrm{cm}$, which is $r_\mathrm{sh}$ at 700~days, assumed.} \\
    \multicolumn{9}{l}{$^b$ $R_\mathrm{CSM}=3.3\times 10^{16}~\mathrm{cm}$, which is $r_\mathrm{sh}$ at 700~days, assumed.}    
	\end{tabular} 
\end{table*}

\section{Comparison with observations}\label{sec:comparison}
In this section, we take two SNe~IIn with long rise times of more than 100~days (HSC16aayt and SN~2008iy) as examples and construct their light curve models assuming that the transitional time $t_t$ determines the luminosity peak. By numerically solving Eq.~(\ref{eq:momentum}), we searched for a parameter combination that can reproduce the overall luminosity evolution of HSC16aayt \citep{moriya2019,moriya2021} and SN~2008iy \citep{miller2010}. In order to compare theoretical and observational light curves, we take the observed optical light curves and assume the zero bolometric correction, which is a reasonable approximation in SNe~IIn \citep[e.g.,][]{ofek2014}. In SN~2008iy, the first observed epoch reported in \citet{miller2010} has a similar brightness to the second observed epoch which is about 150~days later. Thus, we assume that the first observed epoch might actually be a precursor that is often observed in SNe~IIn \citep[e.g.,][]{ofek2013,margutti2014,strotjohann2021}.

\begin{figure}
	\includegraphics[width=\columnwidth]{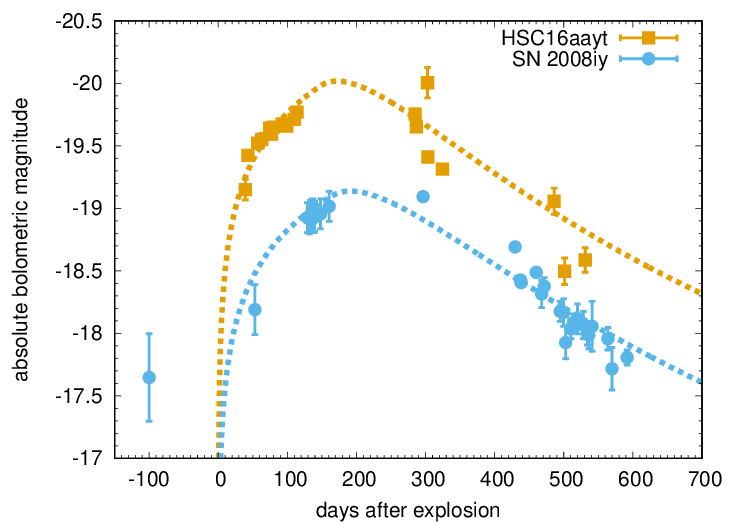}
    \caption{
    Observed and theoretical light curves of HSC16aayt and SN~2008iy. The model parameters for the theoretical models are shown in Table~\ref{tab:models}.
    }
    \label{fig:long}
\end{figure}

Figure~\ref{fig:long} shows comparisons between theoretical and observed light curves. The assumed physical parameters in the theoretical light curves are summarized in Table~\ref{tab:models}. The light curve models for both HSC16aayt and SN~2008iy have a constant CSM density ($s=0$). The shock radius $r_\mathrm{sh}$ reaches $2.4\times 10^{16}~\mathrm{cm}$ (HSC16aayt) and $3.3\times 10^{16}~\mathrm{cm}$ (SN~2008iy) at 700~days, and we provide the CSM mass and diffusion time within these radii in Table~\ref{tab:models}. The transitional times $t_t$ in our models are much longer than $t_\mathrm{diff}$. Thus, the rise times of the two light curve models are determined by $t_t$, and the required CSM mass to accomplish the long rise time ($29-8~\Msun$) is not as massive as 100~\Msun. The estimated properties for SN~2008iy are similar to those in a previous study by \citet{chugai2021}.

If we assume a wind velocity of 100~\kmps, the mass-loss enhancement should last at least for 76~years (HSC16aayt) and 104~years (SN~2008iy), and the average mass-loss rates to form the estimated CSM in HSC16aayt and SN~2008iy are 0.4~\Msunpyr\ and 0.08~\Msunpyr, respectively. The mass loss should be non-steady to form the flat density structure.

\section{Summary}\label{sec:summary}
We have shown that the long rise times ($\gtrsim 100~\mathrm{days}$) observed in some interacting SNe may be caused by a flat CSM density structure ($s\lesssim 1.5$). In such a case, the rise time is determined by the timescale in which the outer SN ejecta with a steep density structure ($\rho_\mathrm{ej}\propto r^{-n}$) are shocked by the reverse shock, not by the diffusion time in the CSM. The CSM with $t_\mathrm{diff}\gtrsim 100~\mathrm{days}$ often has more than 100~\Msun, but the CSM with a flat density structure that has $t_t>t_\mathrm{diff}$ can realize a long rise time with the CSM mass of the order of $10~\Msun$ or less. Therefore, interacting SNe with long rise times may result from the interaction between SN ejecta and flat ($s\lesssim 1.5$) CSM. 

The mass-loss mechanism that leads to the flat density structure is unclear. The mass ejection followed by thermonuclear energy release at the bottom of a hydrogen-rich envelope leads to a CSM density structure of $s\simeq 1.5$ \citep{tsuna2021}, which is still not flat enough to make $\alpha$ positive. Eruptive mass loss caused by, e.g., pulsational pair-instability \citep[e.g.,][]{woosley2007} or luminous blue variables \citep[e.g.,][]{smith2006}, might be relevant to the formation of the CSM with a flat density structure. Photoionization-confined shells are also likely to have a flat density structure \citep{mackey2014}. Because progenitors of interacting SNe with long rise times are likely to experience non-steady mass loss, they can be a valuable probe of mysterious non-steady mass loss in massive stars.

\section*{Acknowledgements}
TJM is supported by the Grants-in-Aid for Scientific Research of the Japan Society for the Promotion of Science (JP20H00174, JP21K13966, JP21H04997).

\section*{Data Availability}
The data underlying this article will be shared on reasonable request to the corresponding author.



\bibliographystyle{mnras}
\bibliography{mnras} 






\bsp	
\label{lastpage}
\end{document}